\documentclass[pra,twocolumn,superscriptaddress,amssymb]{revtex4}
\usepackage{graphicx}
\usepackage{bm}
\usepackage{amsmath,amssymb,fancybox}
\usepackage{color}

\begin{document}
\title{Post hoc verification with a single prover}
\author{Tomoyuki Morimae}
\email{morimae@gunma-u.ac.jp}
\affiliation{ASRLD Unit, Gunma University, 1-5-1 Tenjin-cho Kiryu-shi
Gunma-ken, 376-0052, Japan}
\author{Joseph F. Fitzsimons}
\email{joseph_fitzsimons@sutd.edu.sg}
\affiliation{Singapore University of Technology and Design,
8 Somapah Road, Singapore 487372}
\affiliation{Centre for Quantum Technologies, National
University of Singapore, 3 Science Drive 2, Singapore 117543}

\begin{abstract}
We propose a simple protocol for the verification of quantum computation after the computation has been performed. Our construction can be seen as an improvement on previous results in that it requires only a single prover, who is restricted to measuring qubits in the $X$ or $Z$ basis, while requiring only one way communication, from the prover to the verifier. We also show similar constant round protocols with purely classical verifiers are not possible, unless BQP is contained in the third level of the polynomial hierarchy.
\end{abstract}
\maketitle  

As quantum technologies begin to push against the frontier of what is computationally feasible to simulate using conventional computing technologies, the question of whether it is possible to verify a quantum computation performed on untrusted hardware has become increasingly important~\cite{aharonov2012quantum}. This task can be naturally cast in terms of a two party scenario: We take Alice to be a user with limited quantum capabilities, who wishes to delegate a quantum computation to be performed by Bob, who has access to a universal quantum computer. However, Alice does not trust Bob to faithfully perform the computation. In this case, if Alice is to make use of the results obtained via Bob, she requires some method of verifying that he has performed the computation as directed.

In recent years, significant progress has been made in addressing this problem of verification. In the ideal case, Alice would be able to verify Bob's compliance through an entirely classical protocol, freeing her from the requirement of possessing some quantum capability. Such schemes, however, have thus far proven elusive. Current approaches to this problem take one of two approaches, either allowing Alice some limited quantum capability, such as the ability to prepare~\cite{fitzsimons2012unconditionally,broadbent2015verify} or measure~\cite{morimae2014verification,hayashi2015verifiable} single qubits, or the possession of a constant sized quantum computer~\cite{aharonov2008interactive}, or adding additional entangled provers to the protocol~\cite{reichardt2013classical,reichardt2013classical2,mckague2013interactive,gheorghiu2015robustness,hajduvsek2015device,hayashi2016self}. 

Across both approaches, blind quantum computation~\cite{broadbent2009universal} has proven itself to be an important building block. Blind quantum computing is a secure quantum computing protocol where a technologically limited Alice can delegate her quantum computing to Bob without leaking the information
about her quantum computing. While not all verification protocols have been constructed from blind quantum computing protocols, almost all have proven to be blind even if this was an unintended consequence of the construction~\cite{aharonov2008interactive}, with the notable exception of Ref.~\cite{fitzsimons2015post}. Within the paradigm of blind computation, two distinct approaches have emerged based on the different quantum capabilities assigned to Alice. In the first case, Alice is taken to have the ability to create single qubit states, which can be sent to Bob for processing~\cite{broadbent2009universal}. In this setting, verification of the delegated computation can be achieved by inserting isolated trap qubits into the computation~\cite{fitzsimons2012unconditionally}. In the second approach, Alice is taken to have the ability to perform single qubit measurements, rather than state preparations~\cite{morimae2013blind}. In this setting, verification can be achieved through the use of stabilizer measurements on qubits transmitted from Bob to Alice~\cite{hayashi2015verifiable}. The relatively low technological requirements for Alice's system have already led to experimental demonstrations of both models of blind computation~\cite{barz2012demonstration,greganti2016demonstration}, as well as trap based verification~\cite{barz2013experimental}.  

Up until very recently, all known verification protocols required interaction during the phase when the computation is performed in order to verify its correctness, effectively hiding the computation. Recently, however, a new approach has emerged in which the verification can be postponed until after the computation has been performed~\cite{fitzsimons2015post}, based on the use of efficient interactive proofs with multiple entangled provers for the local Hamiltonian problem~\cite{fitzsimons2015multiprover,ji2015classical} to verify a witness state which certifies the correctness of the computation. In this paper we make use of a similar witness state approach to construct a single prover interactive proof for BQP which is also post hoc, in the sense that it can be performed after the computation to determine the input has concluded. In the protocol we introduce, the verifier needs only to make measurements of qubits in the $X$ and $Z$ basis. Our result differs significantly from the previous post hoc protocols introduced in~\cite{fitzsimons2015post} in that it requires only a single prover, and requires communication in only one direction. Unlike the approach taken in Ref.~\cite{fitzsimons2015post}, however, we show that it is not possible to ``dequantize'' the role of the verifier: A single prover post hoc verification with a classical verifier and a constant number of classical communications is impossible unless BQP is contained in the third level of the polynomial hierarchy, a widely accepted conjecture in computational complexity theory.

We will restrict our focus to verifying the outcome of decision problems. Let $L$ be a language in BQP~\footnote{As is shown in Ref.~\cite{fitzsimons2015post}, we can consider
more general problems than decision problems, but
for simplicity, here we focus on the latter since more general problems can be reduced to the decision 
problem by considering a modified circuit~\cite{fitzsimons2015post}.}. For an instance $x$,
Alice wants to know whether $x\in L$ or $x\notin L$. Since $L$ is in BQP, there exists, for any polynomial $r$, 
a uniformly generated polynomial size quantum circuit $V_x$ acting on $n=\text{poly}(|x|)$ qubits such that
\begin{itemize}
\item
if $x\in L$ then $P_{V_x}(1)\ge 1-2^{-r(|x|)}$,
\item
if $x\notin L$ then $P_{V_x}(1)\le 2^{-r(|x|)}$,
\end{itemize}
where $P_{V_x}(1)$ is the probability of obtaining 1 when the first qubit of $V_x|0\rangle^{\otimes n}$ is measured in the computational basis.

In the setting we consider, where Alice is capable of measuring individual qubits in the $X$ or $Z$ basis but has no further quantum capability, acting alone Alice is limited to deciding questions in BPP. Thus, in order to decide whether $x$ is in $L$, she needs to make use of Bob's quantum capabilities \footnote{Unless BPP=BQP, a relationship widely deemed extremely unlikely.}. Alice can simply ask Bob to decide whether $x\in L$ or $x\notin L$, but is left with the task of verifying whether Bob's response is in fact correct. 

Let us first consider the case when 
Bob tells Alice $x\in L$ and wants to make Alice believe
the fact.
If Bob is honest, $x$ is really in $L$, but if Bob is dishonest
and trying to fool Alice,
$x$ is not in $L$.

Since $L$ is in BQP, it is also in QMA
with the trivial witness state $|0\rangle^{\otimes w}$
and the verifying circuit $W_x\equiv I^{\otimes w}\otimes V_x$
acting on $|0\rangle^{\otimes w+n}$.
Hence there exists a local Hamiltonian $H$
corresponding to the circuit $W_x$
such that
\begin{itemize}
\item
if $x\in L$ then the ground energy of $H$ is
$\le a$,
\item
if $x\notin L$ then the ground energy of $H$ is $\ge b$,
\end{itemize}
where $b-a\ge\frac{1}{\text{poly}(|x|)}$~\cite{kempe2006complexity}. It is known that $H$ can be a 2-local Hamiltonian with only $X$ and $Z$ operators~\cite{biamonte2008realizable}.

In order to justify his claim that $x\in L$, Bob also sends a state $\rho$ to Alice.
If Bob is honest, it is
a ground state of the Hamiltonian $H$. If we follow a similar construction as in Ref.~\cite{fitzsimons2015post} for the Hamiltonian, then this state will encode the history of the circuit via a Feynman-Kitaev clock~\cite{feynman1986quantum,kitaev2002classical}.
Since the witness state is the trivial state $|0\rangle^{\otimes w}$,
Bob can generate such a history state in a polynomial time.
If Bob is dishonest, $\rho$ can be any state.
Let us write the 2-local Hamiltonian
as $H=\sum_Sd_SS$, where $d_S$ is a real number and
$S$ is a tensor product of Pauli operators
where only two operators are $Z$ or $X$,
and others are $I$.
We define the rescaled Hamiltonian
\begin{eqnarray*}
	H''=\frac{1}{\sum_S2|d_S|}H'=\sum_S\pi_SP_S,
\end{eqnarray*}
where
\begin{eqnarray*}
\pi_S=\frac{|d_S|}{\sum_S|d_S|}\ge0,
\end{eqnarray*}
\begin{eqnarray*}
	H'&=&H+\sum_S|d_S|\\
	  &=&\sum_S|d_S|(I+\text{sign}(d_S)S)\\
	  &=&\sum_S2|d_S|P_S,
\end{eqnarray*}
and
\begin{eqnarray*}
	P_S=\frac{I+\text{sign}(d_S)S}{2}.
\end{eqnarray*}

In order to verify the witness, and hence the computation, Alice randomly chooses $S$ with probability $\pi_S$,
and measures $S$. By this we mean that Alice performs single qubit measurements of only two qubits of $\rho$ in the $X$ or $Z$ basis and computes the product of the measurement results,
discarding the other qubits of $\rho$ without measuring them. Note that this does not require Alice to have a large quantum memory, as she can receive the qubits one at a time, resetting her system between reception of qubits. If she obtains the result $-\text{sign}(d_S)$, she accepts.  It was shown in Ref.~\cite{morimae2015quantum}
that the acceptance probability of such a procedure is
\begin{eqnarray*}
	p_{acc}=1-\frac{1}{\sum_S2|d_S|}
	\big(\mbox{Tr}(H\rho)+\sum_S|d_S|\big),
\end{eqnarray*}
which is
\begin{eqnarray*}
p_{acc}\ge\frac{1}{2}-\frac{a}{\sum_S2|d_S|}
\end{eqnarray*}
when $x\in L$, and
\begin{eqnarray*}
p_{acc}\le\frac{1}{2}-\frac{b}{\sum_S2|d_S|}
\end{eqnarray*}
when $x\notin L$.
Their difference is $1/\text{poly}(|x|)$, and therefore Alice can 
almost distinguish the case where $x\in L$ from the case where
$x\notin L$ with probability of error bounded to be exponentially small with only polynomially
many repetitions. Thus Alice can make use of measurements on the witness state to ensure with arbitrarily high probability of correctness that it is in fact true that $x\in L$, as claimed by Bob.

Let us next consider the case when 
Bob tells Alice $x\notin L$ and wants to prove that fact.
If Bob is honest, $x$ is really not in $L$, but if Bob is dishonest
and trying to fool Alice,
$x$ is in $L$.

Let us define $V_x'\equiv (I^{\otimes n-1}\otimes X)V_x$.
Then, because $L$ is in BQP, 
\begin{itemize}
	\item
		if $x\in L$ then $P_{V_x'}(1)\le 2^{-r(|x|)}$,
	\item
		if $x\notin L$ then $P_{V_x'}(1)\ge 1-2^{-r(|x|)}$.
\end{itemize}
Therefore, there exists a local Hamiltonian $H'$
corresponding to
$W_x'\equiv I^{\otimes w}\otimes V_x'$
such that
\begin{itemize}
\item
if $x\in L$ then the ground energy of $H'$ is
$\ge b$,
\item
if $x\notin L$ then the ground energy of $H'$ is $\le a$.
\end{itemize}
This observation is a trivial consequence of the fact that BQP is closed under complement. Running through the same procedure as in the case where $x\in L$, it then follows that the probability of Alice accepting the witness if $x\in L$ can be made exponentially small.

In the above protocol, Alice measures only two qubits of the state that Bob has sent. In principle, however, this does not need to be the case, as Alice is free to measure each incoming qubit. While we do not pursue this approach in the current manuscript, we note that it may be possible to significantly improve the performance of the verification protocol to better estimate the energy of the witness state. This could occur either by making use of the fact that by taking the products of different subsets of the results of local Pauli measurements can be used to infer the results for up to $\frac{3}{8}$ of all Pauli terms in the Hamiltonian, or by making use of a modified Hamiltonian in order to leverage gap amplification results \cite{aharonov2009detectability}.

One defining feature of post hoc verification is that the number of rounds of communication required to verify the computation does not depend on the length of the computation itself. We conclude by showing that verification protocols with classical verifiers cannot have this property, unless there is an unexpected collapse in computational complexity. 

We proceed by means of contradiction. Let us assume that verification can be achieved
with a single prover and a classical verifier,
with $k$ rounds of communication between them, for some constant $k$, such that the probability of Alice accepting an incorrect output of the computation is at most some constant $c$ bounded below $\frac{1}{2}$.
This in particular means that for any language $L\in{\rm BQP}$, if $x\in L$,
Bob can persuade Alice of this fact, whereas if $x \notin L$ Alice cannot be persuaded. In other words,
Bob and Alice exchange $k$ classical messages and 
\begin{itemize}
	\item
		if $x\in L$ there exists a strategy for Bob such
		that Alice accepts with probability $\ge 1-c$,
	\item
		if $x\notin L$ for any strategy of Bob, Alice accepts with probability $\le c$.
\end{itemize}
This is entirely equivalent to the statement that ${\rm BQP}\subseteq{\rm IP}[k]$.
However, if we combine this with the known results~\cite{goldwasser1986private,arora2009computational}
\begin{eqnarray*}
{\rm IP}[k]\subseteq{\rm AM}
[k+2]={\rm AM}[2]\subseteq\Sigma_3^p,
\end{eqnarray*}
we have to conclude that BQP is contained in the third level of the polynomial hierarchy. Thus, by contradiction, we must conclude that such verification protocols cannot exist unless BQP $\subseteq\Sigma_3^p$. Such an inclusion is considered extremely unlikely given the established relationships between BQP and the entire polynomial hierarchy~\cite{aaronson2010bqp}. 

Note that although the above rules out constant round verification of quantum computation with a purely classical verifier, it only holds in the case where there is only a single prover. In the case where multiple provers are allowed, a protocol for verification exists as shown in Ref.~\cite{fitzsimons2015multiprover}. Furthermore, this no go result cannot be directly extended to the case of single prover protocols with polynomially many rounds, since BQP is in ${\rm PSPACE}={\rm IP}$~\cite{bernstein1997quantum}. This does not, however, imply that such protocols must exist, since the only known ways to construct such a proof make use of a prover with power to decide languages believed to be outside of BQP. Thus the question of whether a purely classical verifier can verify a single quantum prover remains an important open problem.

\acknowledgements
TM is supported by the
Grant-in-Aid for Scientific Research on Innovative Areas
No.15H00850 of MEXT Japan, and the Grant-in-Aid
for Young Scientists (B) No.26730003 of JSPS. JF acknowledges support from Singapore's Ministry of Education and National Research Foundation, and the Air Force Office of Scientific Research under AOARD grant FA2386-15-1-4082. This material is based on research funded in part by the Singapore National Research Foundation under NRF Award NRF-NRFF2013-01.

\bibliographystyle{apsrev}
\bibliography{posthoc}

\begin{thebibliography}{30}
\expandafter\ifx\csname natexlab\endcsname\relax\def\natexlab#1{#1}\fi
\expandafter\ifx\csname bibnamefont\endcsname\relax
  \def\bibnamefont#1{#1}\fi
\expandafter\ifx\csname bibfnamefont\endcsname\relax
  \def\bibfnamefont#1{#1}\fi
\expandafter\ifx\csname citenamefont\endcsname\relax
  \def\citenamefont#1{#1}\fi
\expandafter\ifx\csname url\endcsname\relax
  \def\url#1{\texttt{#1}}\fi
\expandafter\ifx\csname urlprefix\endcsname\relax\def\urlprefix{URL }\fi
\providecommand{\bibinfo}[2]{#2}
\providecommand{\eprint}[2][]{\url{#2}}

\bibitem[{\citenamefont{Aharonov and Vazirani}(2012)}]{aharonov2012quantum}
\bibinfo{author}{\bibfnamefont{D.}~\bibnamefont{Aharonov}} \bibnamefont{and}
  \bibinfo{author}{\bibfnamefont{U.}~\bibnamefont{Vazirani}},
  \bibinfo{journal}{arXiv preprint arXiv:1206.3686}  (\bibinfo{year}{2012}).

\bibitem[{\citenamefont{Fitzsimons and
  Kashefi}(2012)}]{fitzsimons2012unconditionally}
\bibinfo{author}{\bibfnamefont{J.~F.} \bibnamefont{Fitzsimons}}
  \bibnamefont{and} \bibinfo{author}{\bibfnamefont{E.}~\bibnamefont{Kashefi}},
  \bibinfo{journal}{arXiv preprint arXiv:1203.5217}  (\bibinfo{year}{2012}).

\bibitem[{\citenamefont{Broadbent}(2015)}]{broadbent2015verify}
\bibinfo{author}{\bibfnamefont{A.}~\bibnamefont{Broadbent}},
  \bibinfo{journal}{arXiv preprint arXiv:1509.09180}  (\bibinfo{year}{2015}).

\bibitem[{\citenamefont{Morimae}(2014)}]{morimae2014verification}
\bibinfo{author}{\bibfnamefont{T.}~\bibnamefont{Morimae}},
  \bibinfo{journal}{Physical Review A} \textbf{\bibinfo{volume}{89}},
  \bibinfo{pages}{060302} (\bibinfo{year}{2014}).

\bibitem[{\citenamefont{Hayashi and Morimae}(2015)}]{hayashi2015verifiable}
\bibinfo{author}{\bibfnamefont{M.}~\bibnamefont{Hayashi}} \bibnamefont{and}
  \bibinfo{author}{\bibfnamefont{T.}~\bibnamefont{Morimae}},
  \bibinfo{journal}{Physical Review Letters} \textbf{\bibinfo{volume}{115}},
  \bibinfo{pages}{220502} (\bibinfo{year}{2015}).

\bibitem[{\citenamefont{Aharonov et~al.}(2008)\citenamefont{Aharonov, Ben-Or,
  and Eban}}]{aharonov2008interactive}
\bibinfo{author}{\bibfnamefont{D.}~\bibnamefont{Aharonov}},
  \bibinfo{author}{\bibfnamefont{M.}~\bibnamefont{Ben-Or}}, \bibnamefont{and}
  \bibinfo{author}{\bibfnamefont{E.}~\bibnamefont{Eban}},
  \bibinfo{journal}{arXiv preprint arXiv:0810.5375}  (\bibinfo{year}{2008}).

\bibitem[{\citenamefont{Reichardt
  et~al.}(2013{\natexlab{a}})\citenamefont{Reichardt, Unger, and
  Vazirani}}]{reichardt2013classical}
\bibinfo{author}{\bibfnamefont{B.~W.} \bibnamefont{Reichardt}},
  \bibinfo{author}{\bibfnamefont{F.}~\bibnamefont{Unger}}, \bibnamefont{and}
  \bibinfo{author}{\bibfnamefont{U.}~\bibnamefont{Vazirani}},
  \bibinfo{journal}{Nature} \textbf{\bibinfo{volume}{496}},
  \bibinfo{pages}{456} (\bibinfo{year}{2013}{\natexlab{a}}).

\bibitem[{\citenamefont{Reichardt
  et~al.}(2013{\natexlab{b}})\citenamefont{Reichardt, Unger, and
  Vazirani}}]{reichardt2013classical2}
\bibinfo{author}{\bibfnamefont{B.~W.} \bibnamefont{Reichardt}},
  \bibinfo{author}{\bibfnamefont{F.}~\bibnamefont{Unger}}, \bibnamefont{and}
  \bibinfo{author}{\bibfnamefont{U.}~\bibnamefont{Vazirani}}, in
  \emph{\bibinfo{booktitle}{Proceedings of the 4th conference on Innovations in
  Theoretical Computer Science}} (\bibinfo{organization}{ACM},
  \bibinfo{year}{2013}{\natexlab{b}}), pp. \bibinfo{pages}{321--322}.

\bibitem[{\citenamefont{McKague}(2013)}]{mckague2013interactive}
\bibinfo{author}{\bibfnamefont{M.}~\bibnamefont{McKague}},
  \bibinfo{journal}{arXiv preprint arXiv:1309.5675}  (\bibinfo{year}{2013}).

\bibitem[{\citenamefont{Gheorghiu et~al.}(2015)\citenamefont{Gheorghiu,
  Kashefi, and Wallden}}]{gheorghiu2015robustness}
\bibinfo{author}{\bibfnamefont{A.}~\bibnamefont{Gheorghiu}},
  \bibinfo{author}{\bibfnamefont{E.}~\bibnamefont{Kashefi}}, \bibnamefont{and}
  \bibinfo{author}{\bibfnamefont{P.}~\bibnamefont{Wallden}},
  \bibinfo{journal}{New Journal of Physics} \textbf{\bibinfo{volume}{17}},
  \bibinfo{pages}{083040} (\bibinfo{year}{2015}).

\bibitem[{\citenamefont{Hajdu{\v{s}}ek
  et~al.}(2015)\citenamefont{Hajdu{\v{s}}ek, P{\'e}rez-Delgado, and
  Fitzsimons}}]{hajduvsek2015device}
\bibinfo{author}{\bibfnamefont{M.}~\bibnamefont{Hajdu{\v{s}}ek}},
  \bibinfo{author}{\bibfnamefont{C.~A.} \bibnamefont{P{\'e}rez-Delgado}},
  \bibnamefont{and} \bibinfo{author}{\bibfnamefont{J.~F.}
  \bibnamefont{Fitzsimons}}, \bibinfo{journal}{arXiv preprint arXiv:1502.02563}
   (\bibinfo{year}{2015}).

\bibitem[{\citenamefont{Hayashi and Hajdusek}(2016)}]{hayashi2016self}
\bibinfo{author}{\bibfnamefont{M.}~\bibnamefont{Hayashi}} \bibnamefont{and}
  \bibinfo{author}{\bibfnamefont{M.}~\bibnamefont{Hajdusek}},
  \bibinfo{journal}{arXiv preprint arXiv:1603.02195}  (\bibinfo{year}{2016}).

\bibitem[{\citenamefont{Broadbent et~al.}(2009)\citenamefont{Broadbent,
  Fitzsimons, and Kashefi}}]{broadbent2009universal}
\bibinfo{author}{\bibfnamefont{A.}~\bibnamefont{Broadbent}},
  \bibinfo{author}{\bibfnamefont{J.}~\bibnamefont{Fitzsimons}},
  \bibnamefont{and} \bibinfo{author}{\bibfnamefont{E.}~\bibnamefont{Kashefi}},
  in \emph{\bibinfo{booktitle}{Foundations of Computer Science, 2009. FOCS'09.
  50th Annual IEEE Symposium on}} (\bibinfo{organization}{IEEE},
  \bibinfo{year}{2009}), pp. \bibinfo{pages}{517--526}.

\bibitem[{\citenamefont{Fitzsimons and
  Hajdu{\v{s}}ek}(2015)}]{fitzsimons2015post}
\bibinfo{author}{\bibfnamefont{J.~F.} \bibnamefont{Fitzsimons}}
  \bibnamefont{and}
  \bibinfo{author}{\bibfnamefont{M.}~\bibnamefont{Hajdu{\v{s}}ek}},
  \bibinfo{journal}{arXiv preprint arXiv:1512.04375}  (\bibinfo{year}{2015}).

\bibitem[{\citenamefont{Morimae and Fujii}(2013)}]{morimae2013blind}
\bibinfo{author}{\bibfnamefont{T.}~\bibnamefont{Morimae}} \bibnamefont{and}
  \bibinfo{author}{\bibfnamefont{K.}~\bibnamefont{Fujii}},
  \bibinfo{journal}{Physical Review A} \textbf{\bibinfo{volume}{87}},
  \bibinfo{pages}{050301} (\bibinfo{year}{2013}).

\bibitem[{\citenamefont{Barz et~al.}(2012)\citenamefont{Barz, Kashefi,
  Broadbent, Fitzsimons, Zeilinger, and Walther}}]{barz2012demonstration}
\bibinfo{author}{\bibfnamefont{S.}~\bibnamefont{Barz}},
  \bibinfo{author}{\bibfnamefont{E.}~\bibnamefont{Kashefi}},
  \bibinfo{author}{\bibfnamefont{A.}~\bibnamefont{Broadbent}},
  \bibinfo{author}{\bibfnamefont{J.~F.} \bibnamefont{Fitzsimons}},
  \bibinfo{author}{\bibfnamefont{A.}~\bibnamefont{Zeilinger}},
  \bibnamefont{and} \bibinfo{author}{\bibfnamefont{P.}~\bibnamefont{Walther}},
  \bibinfo{journal}{Science} \textbf{\bibinfo{volume}{335}},
  \bibinfo{pages}{303} (\bibinfo{year}{2012}).

\bibitem[{\citenamefont{Greganti et~al.}(2016)\citenamefont{Greganti, Roehsner,
  Barz, Morimae, and Walther}}]{greganti2016demonstration}
\bibinfo{author}{\bibfnamefont{C.}~\bibnamefont{Greganti}},
  \bibinfo{author}{\bibfnamefont{M.-C.} \bibnamefont{Roehsner}},
  \bibinfo{author}{\bibfnamefont{S.}~\bibnamefont{Barz}},
  \bibinfo{author}{\bibfnamefont{T.}~\bibnamefont{Morimae}}, \bibnamefont{and}
  \bibinfo{author}{\bibfnamefont{P.}~\bibnamefont{Walther}},
  \bibinfo{journal}{New Journal of Physics} \textbf{\bibinfo{volume}{18}},
  \bibinfo{pages}{013020} (\bibinfo{year}{2016}).

\bibitem[{\citenamefont{Barz et~al.}(2013)\citenamefont{Barz, Fitzsimons,
  Kashefi, and Walther}}]{barz2013experimental}
\bibinfo{author}{\bibfnamefont{S.}~\bibnamefont{Barz}},
  \bibinfo{author}{\bibfnamefont{J.~F.} \bibnamefont{Fitzsimons}},
  \bibinfo{author}{\bibfnamefont{E.}~\bibnamefont{Kashefi}}, \bibnamefont{and}
  \bibinfo{author}{\bibfnamefont{P.}~\bibnamefont{Walther}},
  \bibinfo{journal}{Nature Physics} \textbf{\bibinfo{volume}{9}},
  \bibinfo{pages}{727} (\bibinfo{year}{2013}).

\bibitem[{\citenamefont{Fitzsimons and
  Vidick}(2015)}]{fitzsimons2015multiprover}
\bibinfo{author}{\bibfnamefont{J.}~\bibnamefont{Fitzsimons}} \bibnamefont{and}
  \bibinfo{author}{\bibfnamefont{T.}~\bibnamefont{Vidick}}, in
  \emph{\bibinfo{booktitle}{Proceedings of the 2015 Conference on Innovations
  in Theoretical Computer Science}} (\bibinfo{organization}{ACM},
  \bibinfo{year}{2015}), pp. \bibinfo{pages}{103--112}.

\bibitem[{\citenamefont{Ji}(2015)}]{ji2015classical}
\bibinfo{author}{\bibfnamefont{Z.}~\bibnamefont{Ji}}, \bibinfo{journal}{arXiv
  preprint arXiv:1505.07432}  (\bibinfo{year}{2015}).

\bibitem[{\citenamefont{Kempe et~al.}(2006)\citenamefont{Kempe, Kitaev, and
  Regev}}]{kempe2006complexity}
\bibinfo{author}{\bibfnamefont{J.}~\bibnamefont{Kempe}},
  \bibinfo{author}{\bibfnamefont{A.}~\bibnamefont{Kitaev}}, \bibnamefont{and}
  \bibinfo{author}{\bibfnamefont{O.}~\bibnamefont{Regev}},
  \bibinfo{journal}{SIAM Journal on Computing} \textbf{\bibinfo{volume}{35}},
  \bibinfo{pages}{1070} (\bibinfo{year}{2006}).

\bibitem[{\citenamefont{Biamonte and Love}(2008)}]{biamonte2008realizable}
\bibinfo{author}{\bibfnamefont{J.~D.} \bibnamefont{Biamonte}} \bibnamefont{and}
  \bibinfo{author}{\bibfnamefont{P.~J.} \bibnamefont{Love}},
  \bibinfo{journal}{Physical Review A} \textbf{\bibinfo{volume}{78}},
  \bibinfo{pages}{012352} (\bibinfo{year}{2008}).

\bibitem[{\citenamefont{Feynman}(1986)}]{feynman1986quantum}
\bibinfo{author}{\bibfnamefont{R.~P.} \bibnamefont{Feynman}},
  \bibinfo{journal}{Foundations of physics} \textbf{\bibinfo{volume}{16}},
  \bibinfo{pages}{507} (\bibinfo{year}{1986}).

\bibitem[{\citenamefont{Kitaev et~al.}(2002)\citenamefont{Kitaev, Shen, and
  Vyalyi}}]{kitaev2002classical}
\bibinfo{author}{\bibfnamefont{A.~Y.} \bibnamefont{Kitaev}},
  \bibinfo{author}{\bibfnamefont{A.}~\bibnamefont{Shen}}, \bibnamefont{and}
  \bibinfo{author}{\bibfnamefont{M.~N.} \bibnamefont{Vyalyi}},
  \emph{\bibinfo{title}{Classical and quantum computation}},
  vol.~\bibinfo{volume}{47} (\bibinfo{publisher}{American Mathematical Society
  Providence}, \bibinfo{year}{2002}).

\bibitem[{\citenamefont{Morimae et~al.}(2015)\citenamefont{Morimae, Nagaj, and
  Schuch}}]{morimae2015quantum}
\bibinfo{author}{\bibfnamefont{T.}~\bibnamefont{Morimae}},
  \bibinfo{author}{\bibfnamefont{D.}~\bibnamefont{Nagaj}}, \bibnamefont{and}
  \bibinfo{author}{\bibfnamefont{N.}~\bibnamefont{Schuch}},
  \bibinfo{journal}{arXiv preprint arXiv:1510.06789}  (\bibinfo{year}{2015}).

\bibitem[{\citenamefont{Aharonov et~al.}(2009)\citenamefont{Aharonov, Arad,
  Landau, and Vazirani}}]{aharonov2009detectability}
\bibinfo{author}{\bibfnamefont{D.}~\bibnamefont{Aharonov}},
  \bibinfo{author}{\bibfnamefont{I.}~\bibnamefont{Arad}},
  \bibinfo{author}{\bibfnamefont{Z.}~\bibnamefont{Landau}}, \bibnamefont{and}
  \bibinfo{author}{\bibfnamefont{U.}~\bibnamefont{Vazirani}}, in
  \emph{\bibinfo{booktitle}{Proceedings of the forty-first annual ACM symposium
  on Theory of computing}} (\bibinfo{organization}{ACM}, \bibinfo{year}{2009}),
  pp. \bibinfo{pages}{417--426}.

\bibitem[{\citenamefont{Goldwasser and Sipser}(1986)}]{goldwasser1986private}
\bibinfo{author}{\bibfnamefont{S.}~\bibnamefont{Goldwasser}} \bibnamefont{and}
  \bibinfo{author}{\bibfnamefont{M.}~\bibnamefont{Sipser}}, in
  \emph{\bibinfo{booktitle}{Proceedings of the eighteenth annual ACM symposium
  on Theory of computing}} (\bibinfo{organization}{ACM}, \bibinfo{year}{1986}),
  pp. \bibinfo{pages}{59--68}.

\bibitem[{\citenamefont{Arora and Barak}(2009)}]{arora2009computational}
\bibinfo{author}{\bibfnamefont{S.}~\bibnamefont{Arora}} \bibnamefont{and}
  \bibinfo{author}{\bibfnamefont{B.}~\bibnamefont{Barak}},
  \emph{\bibinfo{title}{Computational complexity: a modern approach}}
  (\bibinfo{publisher}{Cambridge University Press}, \bibinfo{year}{2009}).

\bibitem[{\citenamefont{Aaronson}(2010)}]{aaronson2010bqp}
\bibinfo{author}{\bibfnamefont{S.}~\bibnamefont{Aaronson}}, in
  \emph{\bibinfo{booktitle}{Proceedings of the forty-second ACM symposium on
  Theory of computing}} (\bibinfo{organization}{ACM}, \bibinfo{year}{2010}),
  pp. \bibinfo{pages}{141--150}.

\bibitem[{\citenamefont{Bernstein and Vazirani}(1997)}]{bernstein1997quantum}
\bibinfo{author}{\bibfnamefont{E.}~\bibnamefont{Bernstein}} \bibnamefont{and}
  \bibinfo{author}{\bibfnamefont{U.}~\bibnamefont{Vazirani}},
  \bibinfo{journal}{SIAM Journal on Computing} \textbf{\bibinfo{volume}{26}},
  \bibinfo{pages}{1411} (\bibinfo{year}{1997}).

\end{thebibliography}

\end{document}